\documentclass[aps,prd,twocolumn,nofootinbib]{revtex4-2}
\usepackage{epsfig}
\usepackage{graphicx}
\usepackage{dcolumn}
\usepackage{amssymb,amsmath}
\usepackage{mathrsfs}
\begin{document}

\title{Resonance nuclear excitation of the $^{229}$Th nucleus via electronic bridge process in Th~II.}

\author{V. A. Dzuba and V. V. Flambaum}

\affiliation{School of Physics, University of New South Wales, Sydney 2052, Australia}

\begin{abstract}

The 8.4 eV transition in the $^{229}$Th nucleus is the basis for a high-precision nuclear clock with exceptional sensitivity to new physics effects. We have identified several cases in the Th$^+$ ion where electronic excitations closely resonate with the nuclear excitation, with the smallest energy difference being $\Delta = -0.09$ cm$^{-1}$. We investigate the electronic bridge process, in which nuclear excitation is induced via electronic transitions, and demonstrate that a proper selection of laser frequencies can lead to a dramatic enhancement of this effect. Additionally, we show that the interaction with electrons significantly  shortens the lifetime of the nuclear excited state.

  \end{abstract}

\maketitle

\section{Introduction}

%It was suggested in Ref. ~\cite{Th1} that $^{229}$Th  nucleus has a low energy metastable state. Ref.  ~\cite{Th2}  suggested to use transition between the ground and metastable states as the basis for nuclear clock. Precision of this clock in a "stretched" state  with maximal projections of both electronic and nuclear angular momenta is expected to be very high since in such state electron and nuclear variables are not entangled ~\cite{Kuzmich}. Indeed, the nucleus in such state is practically decoupled from  external fields produced, for example, by the black body radiation.This leads to suppression of the corresponding systematic effects. 

It was suggested in Ref.\cite{Th1} that the $^{229}$Th nucleus has a low-energy metastable state. Ref.\cite{Th2} proposed using the transition between the ground and metastable states as the basis for a nuclear clock. The precision of such a clock in a "stretched" state, with maximal projections of both electronic and nuclear angular momenta, is expected to be very high because, in this state, the electron and nuclear variables are not entangled~\cite{Kuzmich}. Indeed, in such a state, the nucleus is practically decoupled from external fields, such as those produced by blackbody radiation. This leads to a significant suppression of corresponding systematic effects.

On the other hand, the frequency of this transition is highly sensitive to effects of physics beyond the Standard Model, including violations of the Lorentz invariance and the Einstein equivalence principle~\cite{Flambaum2016}, as well as variations in fundamental physical constants~\cite{Flambaum}, such as the fine-structure constant $\alpha$, quark masses, and the QCD scale $\Lambda_{QCD}$. This sensitivity arises from the very small transition energy (8.4 eV), which is the result of strong cancellations between contributions from strong and electromagnetic nuclear interactions, typically at the MeV scale. This cancellation makes the transition frequency extremely sensitive to small changes in these contributions caused by variations in fundamental constants. Such variations may result from interactions with scalar or axion dark matter fields. Therefore, measuring these variations in clock experiments may provide an efficient method for searching for these fields~\cite{Arvanitaki,Stadnik}.

%On the other hand, frequency of this transition is very sensitive to effects of physics beyond the standard model including  Lorentz invariance and Einstein equivalence principle violations ~\cite{Flambaum2016} and  variation of fundamental physical constants ~\cite{Flambaum} (such as the fine structure constant $\alpha$, quark masses and QCD scale $\Lambda_{QCD}$).  The enhancement is due to a very small value of the transition frequency (8.4 eV) which is a result  of the strong cancellation  between contributions to this frequency from the strong and electromagnetic    
%interactions which have MeV scale in nuclei. This makes the transition frequency highly sensitive to small changes in each of these contributions resulting from the variation of the fundamental constants. Variation of the fundamental constants may be due interaction with scalar or axion dark matter fields. Therefore, the measurements of such variation in the clock experiments may be an efficient method to search for such fields  ~\cite{Arvanitaki,Stadnik}.

The measurement of the energy of the nuclear clock transition in $^{229}$Th has been an ongoing effort for many years (see, e.g., \cite{wn1,wn2,wn3}). Recent advances in laser spectroscopy methods have led to a breakthrough in accuracy\cite{Th-wn0,Th-wn,Th-wn1}. The transition frequency  has been measured to be \cite{Th-wn}  $\omega_N = 2,020,407,384,335(2)$~kHz~\cite{Th-wn1}  (67393 cm$^{-1}$) in $^{229}$Th atoms embedded in a solid. Even higher accuracy is anticipated in future experiments with $^{229}$Th ions~\cite{Kuzmich}.

%The measurement of the energy of the nuclear clock transition in $^{229}$Th continued  for many years (see. e.g. ~\cite{wn1,wn2,wn3}). The laser spectroscopy methods recently led to a breakthrough in the accuracy ~\cite{Th-wn0,Th-wn,Th-wn1}. The transition energy has been measured to be $\omega_N = 8.355733(2)_{\rm stat}(10)_{\rm sys}$~eV (67393 cm$^{-1}$)~\cite{Th-wn} and $\omega_N = 2,020,407,384,335(2)$~kHz~\cite{Th-wn1} in Th atoms inside solids.  A higher accuracy is expected in experiments with  Th ions ~\cite{Kuzmich}.  

The amplitude of the nuclear M1 transition is suppressed by five orders of magnitude compared to typical atomic E1 transitions. However, it is possible to enhance this transition using the electronic bridge (EB) process, where electronic transitions induce nuclear transitions via magnetic interaction between electron and nucleus.
%hyperfine interaction  (hfi). 
The EB effect also increases the probability of decay of the nuclear excited state~\cite{Strizhov}.
The high efficiency of excitation of the low-lying isomeric level in
the $^{229}$Th nucleus by photons through the EB was predicted about 30 years ago in the works of Refs.~\cite{Tkalya1,Tkalya2,Karpeshin1992}.
More calculations have been performed for Th II, Th III, and Th IV ions, see e.g. ~\cite{EB1,EB2,EB3,EB4,EB5,EB6,EB7,EB8,EB9,EB10}.

%The amplitude of the nuclear M1 transition is suppressed by five orders of magnitude compared to typical atomic E1 transitions. It is possible  to enhance this transition using  the electronic bridge (EB) process, in which electronic transitions induce nuclear transitions via hyperfine interaction. The EB effect also increases nuclear decay probability. Corresponding calculations have been performed for   Th II, Th III and  Th IV ions  ~\cite{EB1,EB2,EB3,EB4,EB5,EB6,EB7,EB8}.

Our previous EB calculations for the decay~\cite{EB2} and excitation~\cite{EB3} of the excited nuclear state in the Th II ion could not provide accurate results due to a large uncertainty in  the measured value of the nuclear excitation energy and the absence of  measurements of the positions of Th II electron energy levels near the nuclear excitation energy. According to theoretical expectations~\cite{LevelDensity}, the spacing between electron energy levels in this region is very small, making resonance with the nuclear excitation possible.

The aim of this paper is to propose an atomic excitation scheme that maximizes the EB effect and to estimate the magnitude of this effect. Additionally, we estimate the impact of the EB process on the decay of the nuclear excited state. Recent data on the Th II electron spectrum~\cite{Thspectra} (see Supplemental Material for the most complete data) allow us to identify several cases where a small energy denominator in the EB excitation formula leads to a very large  resonance enhancement of the EB effect.

 The error in the calculated level positions exceeds the very small spacing between the measured electron energy levels in highly excited states. To estimate the EB effects, we adopt a statistical approach (based on the  calculations of many possible EB transition probabilities) due to the current lack of information needed to match measured energy levels with calculated electron wave functions. Accurate measurements of the magnetic moments and angular momenta of the Th II electron states would assist in the theoretical identification of the measured levels and improve the accuracy of EB predictions.

\section{Electronic bridge for nuclear excitation  in Th~II}

\begin{figure}[tb]
	\epsfig{figure=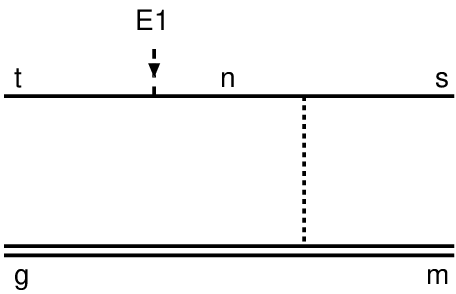,scale=0.6}
	\caption{A diagram for the EB excitation process in Th~II. It is assumed that the atom is in initial electronic state $t$; $n$ and $s$ are intermediate and final electronic states, $g$ ground nuclear state, $m$ is isomeric nuclear state. }
	\label{f:eb}
\end{figure}

We follow the method used in Refs.~\cite{EB3,EB8} for the Th~II anf Th~III ions.  The decay of any atomic state with energy $E_n > \omega_N$ may include nuclear excitation. Since the nuclear excitation energy $\omega_N = 67393$ cm$^{-1}$ lies outside the optical region, we consider a two-step excitation of the electronic states(see also \cite{EB3}). In the first step, the atom is excited from the ground state (GS) to one of the intermediate states $t$ 
used in \cite{EB3}.
In the second step, the ion is further excited by a second laser with frequency which is chosen to satisfy the energy conservation condition for simultaneous nuclear excitation and excitation of the final electronic state, $\omega_1 + \omega_2 = \omega_N + E_s$ ($\omega_1 = E_t$). Nuclear excitation is due to the hyperfine interaction between atomic electrons and the nucleus. Corresponding diagram is presented in Fig.~\ref{f:eb}.
Its calculation involves a summation over all intermediate states $n$. However, in this analysis, we focus on the dominant contributions, which come from specific intermediate states $n$ and final states $s$ providing a very small energy denominator in Eq. (\ref{e:g2a}).
Note that  the final state $s$ is not the ground state but a low-lying excited state satisfying selection rules for the %hfi-induced 
E1 transition  induced by magnetic interaction between electron and nucleus,
from the state $t$ (see diagram in Fig. \ref{f:eb}). 
Suggestion that  EB excitation process with final electron in an excited state has higher probability was made in Ref. ~\cite{Kar2}.

We do not go to highly excited states $s$ to keep laser frequencies within optical region. 
For each final electron state $s$, we look for an intermediate state $n$ (using energy level data from Ref.~\cite{Thspectra}) which corresponds to the smallest energy denominator $\Delta$ in (\ref{e:g2a}), $\Delta = E_n-E_s-\omega_N$, and use this state for estimation of the EB amplitude.
 
The rate of an induced excitation  from state $b$ to state $a$ $W_{ba}^{\rm in}$ can be calculated using the rate of a spontaneous transition $a \rightarrow b$ $W_{ab}$  \cite{Sobelman}:
\begin{equation}\label{e:ab-ba}
 W_{ba}^{\rm in} = W_{ab}\frac{4\pi^3c^2}{\omega^3} I_{\omega}.
 \end{equation}
 Here $I_{\omega}$ is the intensity of isotropic and unpolarized incident radiation. Index $b$ corresponds to nuclear ground state and electron excited state, index $a$ corresponds to nuclear excited state and electronic ground or low energy excited state.
 The rate of a spontaneous transition via electronic bridge process is given by \cite{EB1}
 \begin{equation}\label{e:geb}
 W_{ab} = \frac{4}{9}\left(\frac{\omega}{c}\right)^3 \frac{|\langle I_g||M_k||I_m \rangle|^2}{(2I_m+1)(2J_t+1)}G_2^{(k)},
 \end{equation}
Keeping in mind the relation (\ref{e:ab-ba}), we assume that $\omega$ in (\ref{e:geb}) is the frequency of second excitation ($\omega \equiv \omega_2$). %which is chosen to get into a resonance situation, $\omega_2=\epsilon_s + \omega_N -\omega_1$.
Factor $G_2$ in (\ref{e:geb}) depends on electrons only. It corresponds to upper line of Fig.~\ref{f:eb}. 
%In principle, it has summation over complete set of intermediate states  $n$, see e.g. \cite{EB1}). However, 
Assuming resonance situation and keeping only one strongly dominating term, we have
\begin{equation}\label{e:g2a}
G_2^{(k)} \approx \frac{1}{2J_n+1}\left[\frac{\langle s || T_k || n\rangle \langle n||D|| t \rangle}{\omega_{ns} - \omega_N}\right]^2.
\end{equation}
Here $T_k$ is the electron part of the hyperfine interaction operator (magnetic dipole (M1) for $k=1$ and electric quadrupole (E2) for $k=2$),
$D$ is the electric dipole operator (E1). States $n$ and $s$ are chosen to get close to a resonance, $\epsilon_n-\epsilon_s \equiv \omega_{ns} \approx \omega_N$.

It is convenient to present the results in terms of dimensionless ratios ($\beta$) of EB  transition rates to the nuclear transition rates.
\begin{equation}\label{e:beta}
\beta_{M1}=\Gamma^{(1)}_{\rm EB}/\Gamma_N(M1), \ \  \beta_{E2}=\Gamma^{(2)}_{\rm EB}/\Gamma_N(E2).
\end{equation}
Here $\Gamma^{(k)}$ is given by (\ref{e:geb}) ($\Gamma^{(k)} \equiv W_{ab}$). Both parameters, $\beta_{M1}$ and $\beta_{E2}$ can be expressed via $G2$, (Eq.~(\ref{e:g2a}))~\cite{EB4}
\begin{eqnarray}\label{e:beta1}
&&\beta_{M1}= \left(\frac{\omega}{\omega_N}\right)^3\frac{G_2^{(1)}}{3(2J_s+1)}, \\ 
\label{e:beta2}
&&\beta_{E2}= \left(\frac{\omega}{\omega_N}\right)^3\frac{4G_2^{(2)}}{k_n^2(2J_s+1)}.
\end{eqnarray}
The $M1$ and $E2$ contributions can be combined into one effective parameter $\tilde \beta$ using the known ratio of the widths of the nuclear $M1$ and $E2$ transitions, $\gamma = \Gamma_{\gamma}(E2)/\Gamma_{\gamma}(M1) = 6.9 \times 10^{-10}$, \cite{PRC}.
Then $\tilde \beta = \beta_{M1}(1+\rho)$, where $\rho = \gamma \beta_{E2}/\beta_{M1}$.

\section{Numerical calculations of  the electronic bridge contribution to the  nuclear excitation}

\begin{table*} 
  \caption{\label{t:beta} Versions of the EB process with resonant nuclear excitation. The notations $t$, $n$, and $s$ refer to the diagram in Fig.~\ref{f:eb}.
 It is assumed that $E_t$ = 36390.53 cm$^{-1}$, $\omega_2 = \omega_N +E_s-E_t$ is the frequency of second laser excitation.
  Final states $s$ are listed with their experimental energies ($E_s$) and calculated lifetimes ($\tau_s$).
 The  energy denominator in Eq.~(\ref{e:g2a}) is given by $\Delta = E_n - E_s - \omega_N$.  The third column lists all possible values of $J_n$ from the experimental data in Ref.~\cite{Thspectra}, 
 the last two columns present the calculated values of the EB enhancement factors $\beta$ for specific $J_n$. Here, $x \equiv \langle s || T_k || n \rangle^2 \langle n || D || t \rangle^2$, and $\langle x \rangle$ represents the average value of $x$ over theoretical states that are candidates to represent the actual states $t$ and $n$. No averaging is performed over $s$ since it is a low-lying state that is easily identified in the calculations. The median value of $x$ is denoted by $x_m$. Numbers in square brackets indicate powers of ten.}
 % $x \equiv \langle s || T_k || n\rangle^2 \langle n||D|| t \rangle^2$,
 %$\langle x \rangle$ is the value of $x$ averaged over theoretical states which are candidates to represent actual states $t$ and $n$. 
 %Note that there is no averaging over $s$ since it is low-lying state which is easily identified in the calculations;  
 %$x_m$ is the median  value of $x$.
 %Numbers in square brackets stand for powers of ten. }
 \begin{ruledtabular}
\begin{tabular}{c cc cc cc c c rlc}
$N$ & $E_n$\footnotemark[1] & $J_n$ &  \multicolumn{2}{c}{State $s$} & $E_s$\footnotemark[2] & $J_s$ & $\tau_s$ & $\omega_2$ & 
 \multicolumn{1}{c}{$\Delta$} &
\multicolumn{1}{c}{$\beta(\langle x \rangle)$} &
\multicolumn{1}{c}{$\beta(x_m)$} \\
     &  [cm$^{-1}$] &&&& [cm$^{-1}$] &&& [cm$^{-1}$] & [cm$^{-1}$]  &  & \\ 
\hline
%     En        Es          w         De       beta(<x2>)   beta(x2m)
1 &  73637.54 & 3/2 or 5/2 & $6d^27s$ & $^4$P$_{1/2}$ & 6244.29 & 1/2 & 7.8~s & 37247.10 &  -0.09 &  5.5[6]\footnotemark[3] &  1.6[5]\footnotemark[3] \\  
2 &  73637.54 & 3/2 or 5/2 &  $6d^27s$ & $^4$P$_{1/2}$ &6244.29 & 1/2 &            & 37247.10 &  -0.09 &  1.2[4]\footnotemark[4] &  660\footnotemark[4] \\
3 &  74396.40 & 3/2 or 5/2 & $6d^3$ & $_{3/2}$ &7001.42 & 3/2 & 110~ms & 38004.23 &   1.64 &  2.5[4] &  800 \\  
4 &  76788.94 & 1/2 or 3/2 &  $6d^3$ & $^4$F$_{5/2}$ &9400.96 & 5/2 & 1.5~ms & 40403.77 &  -5.36 &  5.6[4]\footnotemark[5] &  700\footnotemark[5] \\  
5 &  76895.40 &      3/2       &  $6d^3$ & $^4$F$_{5/2}$ &9400.96 & 5/2 &  & 40403.77 &   101.1 &  3.0[3]\footnotemark[6] &  4\footnotemark[6]  \\  

6 &  76445.96 & 1/2 or 3/2 & $6d^27s$ & $^4$P$_{5/2}$ & 9061.10 & 5/2 & 0.7~ms & 40063.91 &  -8.48 &  7.1[3]\footnotemark[5] &  24\footnotemark[5] \\
7 &  76508.84 &  3/2 - 7/2 & $6d^27s$ & $^4$P$_{5/2}$ & 9061.10 & 5/2 &  & 40063.91 &  54.4 &  1.1[3]\footnotemark[7] &  4\footnotemark[7] \\

8 &  68921.30 & 3/2 or 5/2 & $6d^27s$ & $^4$F$_{5/2}$ &1521.90 & 5/2 & 12~s & 32524.71 &   6.06 &  2.6[3] &  28 \\
9 &  75840.96 & 3/2 or 5/2 & $6d^3$ & $_{3/2}$ & 8460.35 & 5/2 & 2.6~ms & 39463.16 & -12.73 &  5.9[3] &  94 \\
\end{tabular}
\footnotetext[1]{Ref.~\cite{Thspectra}, Supplemental Material.}
\footnotetext[2]{	NIST database, \cite{NIST}.}		
\footnotetext[3]{ if $J_n$=3/2 for $E_n$=73637.54 cm$^{-1}$, the  dominant contribution comes from the magnetic dipole  interaction between electron and nucleus.}
\footnotetext[4]{ if $J_n$=5/2 for $E_n$=73637.54 cm$^{-1}$, only electric quadrupole interaction between electron and nucleus contributes.}
\footnotetext[5]{ if $J_n$=3/2.}
\footnotetext[6]{Leading contribution if $J_n$=1/2 for $E_n$=76788.94 cm$^{-1}$.}
\footnotetext[7]{Leading contribution if $J_n$=1/2 for $E_n$=76445.96 cm$^{-1}$.}
\end{ruledtabular}
\end{table*}

%We are not able to match  measured  energy levels for excited  electrons states with energies $E  \gtrsim  \omega_N/2$  to the calculated wave functions  since the uncertainty in the calculated energy levels exceeds the interval between the measured energy levels. Additional problem is uncertainty in the electron angular momenta of the measured excited states. Therefore,  we used the following approach.  At the first step we identified the final electron state $s$  and electron level $n$ which give very small energy denominator $\Delta = E_n-E_s-\omega_N$ in Eq.  (\ref{e:g2a}). The smallest  $\Delta=-0.09$ cm$^{-1}$  only. We also consider 5 other cases with bigger $\Delta$ where we still may expect enhanced EB contributions. 

We are unable to match the measured energy levels of excited electronic states with energies $E \gtrsim \omega_N/2$ to the corresponding calculated wave functions, as the uncertainty in the calculated energy levels exceeds the spacing between the measured levels. An additional issue is the uncertainty in the electron angular momenta of the measured excited states. To address these challenges, we adopted the following approach.

In the first step, we identified the final electronic state $s$ and the intermediate electronic level $n$ that provide very small energy denominators $\Delta = E_n - E_s - \omega_N$ in Eq.~(\ref{e:g2a}). The smallest value of $\Delta$ is $-0.09$ cm$^{-1}$. We also considered five additional cases with larger $\Delta$, where enhanced electronic bridge contributions may still be expected.

To estimate $\beta$, we evaluated the products of the matrix elements  $x \equiv \langle s || T_k || n\rangle^2 \langle n||D|| t \rangle^2$ (see Eq.  (\ref{e:g2a})) .  
The measured value of the final state  level $s$, which has low energy, can be reliably matched with its corresponding theoretical energy level and wave function. However, we do not have definitive theoretical identifications for the initial state $t$ and the intermediate state $n$ (see the diagram in Fig.\ref{f:eb}).

 For each final state $s$, we calculated 217 possible values of $x$, corresponding to different assignments of the experimental levels $t$ and $n$.
 % For each final state $s$,  we calculated 217 values of $x$ corresponding to different identifications of experimental levels  $t$ and $n$.
 The distribution of these $x$ values indicates that there is no strong configuration mixing. Such mixing would result in  eigenstates  made from chaotic superpositions  of all configurations close in energy and smooth distribution of $x$ values.  Instead, the distribution of $x$ consists of a small group of large values that dominate the average $\langle x \rangle$, and a broad distribution of smaller values.  A similar conclusion was reached in Ref.~\cite{Kar1}.
 Thus, we cannot apply statistical theory developed for chaotic eigenstates~\cite{chaos1,chaos2,chaos3,chaos4,chaos5}. 

%  which would  make  eigenstates chaotic superpositions  of all configurations close in energy.  Therefore, we can not apply statistical theory developed for chaotic eigenstates \cite{chaos1,chaos2,chaos3,chaos4,chaos5}. On the contrary, values of $x$ incudes a small group of big values which dominate average  value of $x$ (denoted by $\langle x \rangle$), and a large number of small values which are broadly distributed. 

The large matrix elements arise when the leading configurations in the initial and final states differ by a single orbital, allowing for a strong single-particle transition. The smaller matrix elements are generated by  configuration mixing. To identify a representative value of $x$, we found  the median value $x_m$ (in the middle of $x$ distribution). Note that $x_m$ is significantly smaller than the average $\langle x \rangle$.

% Big matrix elements appear when leading configurations in initial and final states differ by one orbital and may be connected by a large single-particle transition. Small matrix elements are due to configuration mixing.    
 % Big values of $x$ may be due to presence of $7s$ orbital which can produce large single-particle hfi  matrix elements  $\langle s || T_k || n\rangle$ between  dominating configurations in the states $s$ and $t$, and electric dipole matrix elements allowed by approximate selection rules for single-particle  E1 matrix elements $\langle n||D|| t \rangle$   between  dominating configurations in the states $n$ and $t$.   Small values of $x$ are sensitive to the configuration mixing. 
%To find a typical value of $x$ we calculated median value  $x_m$ (in the middle of $x$ distribution) which happened to be much smaller than $\langle x \rangle$.

The results of our calculations for several possible transitions are presented in Table~\ref{t:beta}. Our findings show that the probability of nuclear excitation can, in principle, be enhanced  5 million times. However, the most probable cases correspond to $\beta(x_m)$, presented in the last column of the table. In these cases, the EB enhancement factor $\beta$  may reach  five orders of magnitude when the magnetic dipole hyperfine interaction is allowed (for $J_n = 3/2$). This enhancement can be achieved through proper selection of laser frequencies in a two-step atomic excitation process, followed by nuclear excitation.

However, as it was noted in Ref. \cite{EB10}, a very large value of EB enhancement  factor $\beta$  does not necessarily lead to the  $\beta$ times  enhancement of the excitation rate. The point is that the photon capture resonance  cross section contains total width  in denominator,  which may be dominated by the radiative width $\Gamma_{\gamma}$,  if it is very strongly enhanced by the EB process ($\Gamma_{\gamma}=(1+\beta) \Gamma_N$, where $\Gamma_N \approx 5 \times 10^{-4}$ Hz  is the the decay width of the nuclear isomeric state in bare nucleus - see e.g \cite{Katori}). This effect can be qualitatively described by the factor, which appears in the resonance photon capture cross section
\begin{equation}     
\sigma \sim \lambda^2 \frac{\Gamma_{\gamma}\Gamma }{(\omega-\omega_0)^2 +\Gamma^2/4},
 \end{equation}
where $\lambda$ is the photon wavelength,  $\Gamma= \Gamma_{\gamma} + \Gamma_{Ni} + \Gamma_s+ \Delta_l$, $\Gamma_{Ni}$ is width of the nuclear transition in  $^{229m}$Th$^+$ , $\Gamma_s$ is the width of the final electron state and $\Delta_l$ is the width of the energy distribution of the  photons in the pumping  laser beam (or width of the frequency comb teeth). There may be other line broadening mechanisms. 

In the resonance $\sigma \propto \beta \Gamma_N/( \beta \Gamma_N  +  \Gamma_{Ni}+ \Gamma_s+ \Delta_l)$.  We see that in the limit $\beta \to \infty$ the enhancement in the center of the resonance disappears.  

However, in a realistic situation there is $\beta$ times enhancement of the cross section.
Consider different contributions to the total width $\Gamma$.  Specifically,  for the  final electron states $^4$P$_{1/2}$ with the energy  6244.29  cm$^{-1}$ and $^5$F$_{5/2}$ with the energy  1521.90  cm$^{-1}$ , the radiative width  $\Gamma_s$   is very small, 0.1 Hz, since it is produced by  low frequency M1 and E2 transitions. For other final states in the Table \ref{t:beta}, E1 transitions are possible, and the width is from 10 to 1000 Hz. Note that all lifetimes are strongly affected by the configuration mixing. 

 There is  an experimental  indication that the   lifetime  of the nuclear isomer in $^{229m}$Th$^+$  with electrons in their ground state is smaller than 10 ms \cite{Th+lifetime} (see discussion in the next section), i.e.  in   $^{229m}$Th$^+$  the natural decay width  $\Gamma_{Ni}  > $ 100 Hz.  When electrons are in exited state, $\Gamma_{Ni}$  would hardly be smaller.  If this is the case,  there is an enhancement of the cross section at least up to $\beta=2 \times 10^5$.  
%    For the laser frequency  bandwidth $\Delta=1$ Hz, there is no  

Moreover,  during the frequency scanning procedure  we are not exactly in the resonance,  i.e.  $(\omega-\omega_0)^2 > \Gamma^2/4$.  The width of the  frequency comb teeth $\Delta_l$ may significantly exceed natural width $\Gamma$. Finally, the scanning time is reduced inversely proportional to the width, so effectively one should  consider the product of the cross section times the total width, so $\beta$ in the denominator cancels out.  In this case the EB enhancement factor $\beta$ is important for any large $\beta$. 
%The results of the calculations for a number of possible transitions are presented in Table~\ref{t:beta}. We see that the probability of the nuclear excitation may, in principle,  be enhanced up to 5 millions  times. However, most probable cases  correspond to $\beta(x_m)$, presented in the last column. Here the enhancement may reach five orders of magnitude if the  magnetic dipole hfi is allowed  (for $J_n=3/2 $). This enhancement may be achieved by proper choices of laser frequencies in a two-step process of atomic excitation which is followed by the nuclear excitation. 

\section{Numerical calculation of  the electronic bridge contribution to the decay of the  nuclear excited state}

\begin{figure}[tb]
	\epsfig{figure=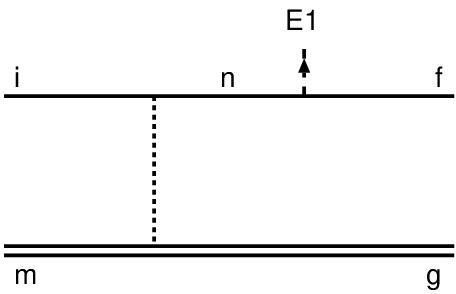,scale=0.6}
	\caption{An EB contribution to the decay rate of the isomeric nuclear state $m$. Index $i$ stands for the atomic ground state, $n$ stands for intermediate states, $f$ stands for final odd-parity states. }
	\label{f:decay}
\end{figure}

\begin{table} 
  \caption{\label{t:decay} Calculations of the coefficient $R$ (\ref{e:R}) at different assumptions for experimental level
at $E_n$=67378~cm$^{-1}$ ($J_n$=3/2,5/2 or 7/2). $N$ is serial number of the state in the calculations for given $J_n$ and parity;
$\Delta E_n$ is the shift in the energy of state $n$ to bring it to experimental energy of 67378~cm$^{-1}$. Numbers in square brackets stand for powers of ten.
}
\begin{ruledtabular}
\begin{tabular}{cc ccc cc}
$N$ & $E_n$~cm$^{-1}$  & $\Delta E_n$ & \multicolumn{3}{c}{Partial $J_f$ contributions} & Sum \\
\hline
 \multicolumn{3}{c}{$J_n$=3/2; $J_f:$}  & 1/2 & 3/2 & 5/2 & \\
 68 &  67601 &    222 &  8.524[1] &  1.414[2] &  8.824[1] &  3.149[2] \\
 69 &  67914 &    536 &  1.908[3] &  3.929[3] &  1.695[3] &  7.532[3] \\
 70 &  68174 &    796 &  4.276[2] &  5.208[2] &  2.036[2] &  1.152[3] \\
 71 &  68568 &   1189 &  6.270[3] &  3.274[3] &  1.933[3] &  1.148[4] \\
 72 &  68866 &   1488 &  2.554[2] &  2.161[2] &  2.261[2] &  6.976[2] \\
 73 &  69131 &   1752 &  6.223[2] &  6.795[2] &  1.279[3] &  2.580[3] \\
 74 &  69391 &   2013 &  1.762[3] &  1.398[3] &  2.280[3] &  5.440[3] \\
 75 &  69500 &   2121 &  1.548[3] &  4.602[3] &  4.080[3] &  1.023[4] \\
 76 &  69719 &   2340 &  8.695[3] &  6.655[3] &  1.121[4] &  2.655[4] \\
 77 &  69874 &   2496 &  5.716[2] &  5.475[2] &  6.200[2] &  1.739[3] \\
 78 &  70221 &   2842 &  5.813[2] &  2.826[3] &  1.475[3] &  4.882[3] \\
 79 &  70583 &   3204 &  2.090[3] &  6.944[3] &  8.203[3] &  1.724[4] \\
 80 &  71018 &   3639 &  1.665[3] &  1.943[3] &  3.999[3] &  7.607[3] \\
 81 &  71104 &   3725 &  2.949[2] &  3.232[2] &  5.333[2] &  1.151[3] \\
 82 &  71270 &   3892 &  3.868[3] &  4.111[3] &  9.680[3] &  1.766[4] \\
 83 &  71528 &   4149 &  1.113[2] &  6.798[1] &  1.302[2] &  3.095[2] \\
 84 &  71816 &   4437 &  1.108[3] &  7.471[2] &  4.730[3] &  6.586[3] \\
 85 &  72150 &   4771 &  2.141[3] &  4.601[3] &  1.353[4] &  2.028[4] \\
 \multicolumn{3}{c}{$J_n$=5/2: $J_f:$} & 3/2 & 5/2 & 7/2 & \\
 78 &  67493 &    115 &  1.624[4] &  2.049[4] &  4.874[3] &  4.160[4] \\
 79 &  67915 &    536 &  9.062[2] &  4.504[2] &  2.893[2] &  1.646[3] \\
 80 &  68350 &    971 &  2.976[4] &  1.724[4] &  2.396[4] &  7.096[4] \\
 81 &  68531 &   1153 &  3.824[4] &  5.077[4] &  4.200[4] &  1.310[5] \\
 82 &  69006 &   1628 &  5.328[2] &  1.447[3] &  7.728[2] &  2.752[3] \\
 83 &  69196 &   1818 &  1.458[4] &  3.942[4] &  3.654[4] &  9.054[4] \\
 84 &  69309 &   1930 &  6.857[3] &  5.275[3] &  2.767[3] &  1.490[4] \\
 85 &  69352 &   1973 &  9.055[2] &  1.242[3] &  6.894[2] &  2.837[3] \\
 86 &  69640 &   2261 &  8.408[3] &  2.196[4] &  1.227[4] &  4.264[4] \\
 87 &  69817 &   2439 &  6.820[3] &  2.231[4] &  1.243[4] &  4.155[4] \\
 88 &  69918 &   2539 &  3.727[2] &  8.864[2] &  6.485[2] &  1.908[3] \\
 89 &  69974 &   2596 &  6.553[2] &  1.908[3] &  2.102[3] &  4.665[3] \\
 90 &  70222 &   2844 &  1.637[3] &  3.367[3] &  2.770[3] &  7.774[3] \\
 91 &  70475 &   3096 &  5.853[3] &  6.302[3] &  3.767[3] &  1.592[4] \\
 92 &  70510 &   3131 &  9.887[3] &  9.563[3] &  5.978[3] &  2.543[4] \\
 93 &  70823 &   3444 &  9.029[1] &  1.072[2] &  6.852[1] &  2.660[2] \\
 94 &  70924 &   3546 &  8.687[2] &  9.090[2] &  5.178[2] &  2.296[3] \\
 95 &  71035 &   3656 &  2.439[2] &  2.602[2] &  1.229[2] &  6.270[2] \\
\end{tabular}			
%\fooM1tnotetext[1]{Only electric quadrupole {\bf magnetic between electron and nucleus} contributes. }
%\footnotetext[2]{Contribution from the next state $E_n$=67509.63 ~cm$^{-1}$, in case of $J_n$=7/2 for $E_n$=67378.61~cm$^{-1}$. }
\end{ruledtabular}
\end{table}

The electronic bridge  contribution to the decay of the nuclear excited state is described by the diagram in Fig.~\ref{f:decay}. This process requires summing over all possible final electronic states $f$. The sum over the intermediate states $n$  is dominated by a single term with energy $E_n = 67378.61$ cm$^{-1}$~\cite{Thspectra}, which results in a small energy denominator in  EB amplitude induced by magnetic dipole or electric quadrupole interaction with the nucleus: $\Delta = E_n - \omega_N = 14.73$ cm$^{-1}$.
Very little is known about this state, only energy and a range of possible values of the total angular momentum ($J_n$=3/2, 5/2 or 7/2). 
Therefore, we consider a number of possible candidates (see below).

As before, it is convenient to present the results in terms of dimensionless ratios of EB  transition rates to the nuclear transition rate.
Note, that summation over final states $f$ leads to summation of terms similar to $\beta_{M1}$ (\ref{e:beta1}) and $\beta_{E2}$ (\ref{e:beta2}) with different values of $\omega$: $\omega \equiv \omega_f = \omega_N - E_f - E_i$. Therefore, we introduce new variable $R$ 
\begin{equation}\label{e:R}
R = \sum_f \beta(\omega_f).
\end{equation}
In the paper ~\cite{Th+lifetime} they study internal conversion signal which  should appear after collision of  the $^{229m}$Th isomer ions with a solid surface.  They see clear signals from Th III and Th IV ions but  do not observe any signal from Th II ions. The authors of Ref.  ~\cite{Th+lifetime} suggested a possible explanation: lifetime  ($\tau$) of the $^{229m}$Th isomer is very small, $\tau < 10~{\rm ms}$, so Th II ions do not reach the surface. 

 The half-life  of the isomeric  nuclear state in  Th IV is practically not affected by the electronic  bridge mechanism \cite{EB1}. This half-life in  Th IV was measured to be $T_{1/2}=1400^{+600}_{-300}$ \cite{Katori}, corresponding to the lifetime $\tau=2020^{+866}_{-433}$. This lifetime  is in a reasonable agreement with the values obtained in the solid state experiments \cite{Th-wn0,Th-wn,Th-wn1,Hiraki,Zhang,Kraemer,Tkalya,Morgan,Perera}  after introducing the solid state corrections.  
To obtain   $\tau < 10~{\rm ms}$,   one needs very large $R>2 \times 10^5$. This small value of the lifetime constitutes the  so called {\em thorium puzzle}.  Possible solutions of this puzzle (very large value of EB coefficient,  $R=56500$, collisional  broadening) were discussed in Refs.  ~\cite{Kar1,Kar2,Kar3}.

We study this problem in present work using new experimental data on Th$^+$ energy levels~\cite{Thspectra}. Strong enhancement of the decay rate due to above mentioned small energy denominator ($\Delta = 14.73$ cm$^{-1}$) may potentially explain the puzzle. 
We perform the calculations summing over large number of intermediate states $n$ around $E_n \sim \omega_N$ (about 20 states for each $J_n$) and all final states with $E_f < \omega_N$ (about 70 states for each $J_f$). 

Comparison with experiment for excitation energies between 0 and 60000~cm$^{-1}$ shows that when approaching the upper limit 60000~cm$^{-1}$,
calculated energies tend to be about 3000~cm$^{-1}$ higher than experiment. Since we need states with energies $E_n \sim \omega_N$ = 67393.34~cm$^{-1}$, it does make sense to shift calculated energies down to make more realistic energy denominators.

The procedure is the following. We choose a particular state in the energy interval 67000~cm$^{-1} < E_n < $72000 ~cm$^{-1}$ and shift its energy to reproduce exactly the experimental energy which is the closest to $\omega_N$, $E_n$=67378.61~cm$^{-1}$~\cite{Thspectra}.
The energies of all other states, included in the summation over $n$, are shifted by the same value. The EB enhancement factor  $R$ is calculated with new energies.
Then the  procedure is repeated for a number of other candidate states in the same energy interval. The results are presented in Table~\ref{t:decay}. 
We only consider states with $J_n$=3/2 and 5/2. We don't consider states with $J_n$=7/2 because they are connected to the ground state, which has $J_i$=3/2, by the electric quadrupole interaction only.
Corresponding matrix elements are significantly smaller than those for magnetic dipole interaction and cannot explain large decay rate.

The largest values of $R$ in the Table, $R = 1.3 \times 10^5$, corresponds to the assumption that the experimental energy  $E_n$=67378.61~cm$^{-1}$  match up with the calculated state number 81 with the calculated energy 68531~cm$^{-1}$ and total angular momentum $J_n=5/2$. This  maximal calculated value of $R$,  $R = 1.3 \times 10^5$, is close to  the experimental lower bound,  $R > 2 \times 10^5$.  Therefore, we cannot entirely exclude this scenario as a potential solution to the thorium puzzle: an unexpectedly large enhancement of the nuclear decay rate via the electronic bridge mechanism~\cite{Kar1,Kar2,Kar3}.

We have also performed calculations for  many  levels located   below nuclear frequency $\omega_N$ = 67393.34~cm$^{-1}$,  but they do not produce any qualitative changes to the picture.

\section{Conclusion}

We have identified several cases of resonance between electronic and nuclear excitations, indicating the potential for significant enhancement of nuclear excitation through the electronic bridge mechanism. This enhancement can be achieved by selecting appropriate laser frequencies in a two-step atomic excitation process, followed by nuclear excitation.

Due to uncertainties in assigning calculated wave functions to the experimentally measured energy levels near the nuclear excitation energy, it was not possible to provide precise enhancement factors. These factors may vary from a thousandfold to a millionfold enhancement. To address this uncertainty, we analyzed hundreds of possible level assignments and presented the average and median values of the enhancement factor $\beta$ for six selected laser frequency configurations exciting both nuclear and electronic transitions (see Table~\ref{t:beta}).

We also estimated the effect of electronic states on the decay lifetime of the nuclear excited state (see Table~\ref{t:decay}). We identified one possible assignment of calculated wave functions and energy levels to observed electron energy levels that is consistent with the current experimental limit on the lifetime of $^{229\text{m}}$Th. However, to explain the unexpectedly short lifetime of the $^{229\text{m}}$Th state in Th~II, it is important to experimentally investigate the possible existence of additional electronic levels near the energy of the nuclear isomeric state.

Accurate measurements of the magnetic moments and angular momenta of the Th II electronic states would help to refine the theoretical identification of the measured levels and improve the accuracy of predictions for the electronic bridge mechanism.

This work was supported by the Australian Research Council Grant No. DP230101058. We are grateful to Feodor Karpeshin for valuable discussions and references.

  \appendix
\section{Method of Calculation}

We use the same computational method as in our previous work on the Th~III ion~\cite{EB8}. The method is based on the relativistic Hartree-Fock (RHF) approach combined with the configuration interaction and single-double coupled cluster (CI+SD) method~\cite{CI+SD} to calculate the three-electron valence states of the Th~II ion.

The calculations begin with the closed-shell ThV ion. The single-electron basis states for the valence electrons are generated in the field of the frozen core using the B-spline technique~\cite{B-spline}. The SD equations are first solved for the core and subsequently for the valence states~\cite{CI+SD}. This procedure results in the creation of one-electron and two-electron correlation operators, $\Sigma_1$ and $\Sigma_2$, which are then applied in the CI calculations.

To compute the transition amplitudes, we employ the time-dependent Hartree-Fock method~\cite{TDHF}, which is equivalent to the well-known random-phase approximation (RPA). Solving the RPA equations for the valence states produces effective operators for the external field, which are used to evaluate matrix elements between the valence states.

It is important to note that the states of interest have excitation energies around 70,000 cm$^{-1}$. For Th~II, this requires calculating approximately one hundred states for each value of $J$ and parity. The density of states in this energy region is high, and the energy intervals between them are smaller than the uncertainty in the calculations.

Consequently, reliably matching the measured energy levels to the calculated levels and wave functions of the high-energy states is currently not feasible. To address this challenge, we employ the averaging procedure described earlier in the manuscript.

\end{document}